\documentclass[fleqn,usenatbib,useAMS]{mnras}

\usepackage{graphicx}	
\usepackage{amsmath}	
\usepackage{amssymb}	
\usepackage{multicol}        
\usepackage{bm}		
\usepackage{pdflscape}	
\usepackage{url}

\newcommand{\targ}{GPM\,J1839\ensuremath{-}10} 

\newcommand{\warwick}{1}
\newcommand{\hamburg}{2}
\newcommand{\sheffield}{3}
\newcommand{\iac}{4}
\newcommand{\sheffieldrse}{5}
\newcommand{\oklahoma}{6}
\newcommand{\colorado}{7}
\newcommand{\cork}{8}

\usepackage[T1]{fontenc}
\usepackage{ae,aecompl}

\usepackage{newtxtext,newtxmath}

\title[Optical counterpart of \targ]{Constraints on an optical counterpart for the long-period radio transient \targ}

\author[Pelisoli et al.]{
Ingrid Pelisoli,$^{\warwick}$\thanks{E-mail: ingrid.pelisoli@warwick.ac.uk}
A.~J. Brown$^{\hamburg}$,
N. Castro Segura$^{\warwick}$,
V.~S. Dhillon$^{\sheffield,\iac}$,
M.~J. Dyer$^{\sheffield,\sheffieldrse}$,
J.~A. Garbutt$^{\sheffield}$,\newauthor
M.~J. Green$^{\oklahoma,\colorado}$
D. Jarvis$^{\sheffield}$,
M. R. Kennedy$^{\cork}$,
P. Kerry$^{\sheffield}$,
S.~P. Littlefair$^{\sheffield}$,
J. McCormac$^{\warwick}$,\newauthor
J. Munday$^{\warwick}$,
S.~G. Parsons$^{\sheffield}$,
E. Pike$^{\sheffield}$,
D.~I. Sahman$^{\sheffield}$,
A.~Yates$^{\sheffield}$
\\
$^{\warwick}$Department of Physics, University of Warwick, Gibbet Hill Road, Coventry, CV4 7AL, UK\\
$^{\hamburg}$Hamburger Sternwarte, University of Hamburg, Gojenbergsweg 112, 21029 Hamburg, Germany\\
$^{\sheffield}$Astrophysics Research Cluster, School of Mathematical and Physical Sciences, University of Sheffield, Sheffield, S3 7RH, United Kingdom\\
$^{\iac}$Instituto de Astrof\'{i}sica de Canarias, E-38205 La Laguna, Tenerife, Spain\\
$^{\sheffieldrse}$Research Software Engineering, University of Sheffield, Sheffield, S1 4DP, United Kingdom\\
$^{\oklahoma}$Homer L. Dodge Department of Physics and Astronomy, University of Oklahoma, 440 W. Brooks Street, Norman, OK 73019, USA\\
$^{\colorado}$JILA, University of Colorado and National Institute of Standards and Technology, 440 UCB, Boulder, CO 80309-0440, USA\\
$^{\cork}$School of Physics, University College Cork, Cork, T12 K8AF, Ireland\\
}
\date{Last updated XXX; in original form XX}

\pubyear{2025}

\begin{document}
\label{firstpage}
\pagerange{\pageref{firstpage}--\pageref{lastpage}}
\maketitle

\begin{abstract}
Long period radio transients (LPTs) are periodic radio sources showing pulsed emission on timescales from minutes to hours. The underlying sources behind this emission are currently unclear. 
There are two leading candidates: neutron stars or white dwarfs. Neutron stars could emit at LPT timescales as magnetars, binaries, or precessing sources. White dwarfs on the other hand have only been observed to emit in radio as binary systems with companions that provide charged particles through their wind. A key distinction is that an optical counterpart is much more likely in the white dwarf scenario. \targ\ is an LPT with a radio period of 21~min for which the white dwarf scenario has been favoured, but no optical counterpart is confirmed. Using HiPERCAM, a high-speed multi-colour photometer that observes simultaneously in $u_sg_sr_si_sz_s$ filters, we probe the existence of a white dwarf in \targ. We do not directly detect a white dwarf, but cannot rule out its presence given the uncertain distance and reddening of \targ. On the other hand, we find evidence in our data for periodic behaviour in harmonics of the radio period, as expected from the white dwarf scenario.
\end{abstract}




\section{Introduction}


Long-period radio transients (LPTs) are a class of systems that show periodic pulsed radio emission with timescales varying from minutes to hours. The first source of the class to be discovered was GCRT~J1745-3009 \citep{Hyman2005}, which showed pulses with a width of 10~min repeating every 77~min. After more than 15 years without new LPTs being reported, there have recently been a number of discoveries \citep{Hurley-Walker2022, Hurley-Walker2023, Caleb2024, Dong2024, Hurley-Walker2024, Li2024, deRuiter2025, Lee2025, Bloot2025} fuelled by radio surveys with larger fields-of-view and better resolution, and by more comprehensive examination of the data considering longer timescales.

Aside from their common characteristic of showing reoccurring radio pulses with periodicities in the range of minutes to hours, LPTs are a highly heterogenous class. Pulses have been observed to disappear or decrease significantly in strength over time \citep[e.g.][]{Hyman2005, Caleb2024}, or to remain active for decades \citep{Hurley-Walker2023}. The duration and profile of the pulses are also highly variable, with some systems showing pulses with complex structure \citep{Hurley-Walker2023} while others present a more smooth profile \citep{deRuiter2025} and some vary between these states \citep{Hurley-Walker2022}. These factors suggest that, rather than having a single common origin, LPTs are explained by different physical mechanisms with similar observational characteristics.

The properties of the emission shown by LPTs (intensity, periodicity, coherence and polarisation) are generally inconsistent with the stellar activity shown by M-dwarfs or brown dwarfs \citep[e.g.][]{Hyman2005}. Instead, a compact object, either a neutron star or a white dwarf, is believed to be the underlying source of emission, though the exact mechanisms at play are unclear. Neutron stars can show pulsed radio emission due to pair-production conditions being met in their magnetic poles, leading to the occurrence of non-thermal emission from accelerated particles that varies periodically as the magnetic poles sweep across the line-of-sight. This, however, normally requires fast ($\lesssim 1$~min) spin periods \citep[e.g.][]{Szary2014}. An alternative is that the emission is powered by the decay of a strong magnetic field rather than fast rotation in what is known as a magnetar \citep{Duncan1992}. X-ray emission is also expected from this mechanism, which is indeed detected from some LPTs \citep[e.g.][]{Li2024}. Other models suggest that the observed long period could be the orbital period of a double neutron star binary \citep{Turolla2005}, or the precession period of the neutron star \citep{Zhu2006}. White dwarfs, on the other hand, are not expected to meet pair production conditions at their typical rotation rates and magnetic fields, with field strengths $\gtrsim 10^9$~G required for efficient pair-production at the usual rotation rates \citep{Rea2024}, contrasting with observed white dwarf field strengths of $\lesssim 10^8$~G \citep[e.g.][]{Hardy2023}. Therefore, an external source of charged particles is required to fuel non-thermal emission, such as a binary companion \citep[e.g][]{Lyutikov2020, Qu2025}.

Corroborating the binary model for radio emission from white dwarf sources, persistent radio emission from isolated white dwarfs has been found to be absent above a level of $1-3$~mJy \citep{Pelisoli2024}, a lower flux level than most LPTs. Although this study relied on the Very Large Array Sky Survey (VLASS) and could thus miss pulsed sources with low duty cycle, all the confirmed white dwarf systems found to show pulsed non-thermal emission are indeed in binary systems with red dwarf companions. This includes the three binary white dwarf pulsars \citep{Marsh2016, Pelisoli2023, CastroSegura2025}, whose pulsed emission is associated with the white dwarf spin, and two LPTs, ILT~J1101+5521 \citep{deRuiter2025} and GLEAM-X~J0704–37 \citep{Hurley-Walker2024, Rodriguez2025}, where the radio periodicity matches the orbital period of the binary. These latter two detections demonstrate that at least some LPTs can be explained as white dwarf plus red dwarf systems, though there are cases where this possibility can be almost certainly excluded \citep{Lyman2025}.

Given their uncertain nature and emission mechanisms, LPTs are currently a puzzle in radio astronomy. With additional surveys such as the Square Kilometre Array Observatory (SKAO) and the Deep Synoptic Array (DSA) 2000 on the horizon, understanding the properties of LPTs is a high priority. The main distinction between the neutron star and white dwarf mechanisms is that white dwarfs are expected to be stronger optical sources. Therefore, optical follow-up, and in particular deep-imaging, is a powerful tool for LPT characterisation. In this work, we report optical observations of \targ, discovered by \citet{Hurley-Walker2023}. \targ\ was found to be active at least since 1988 and shows radio pulses with a period of 1318.1957(2)~s ($\approx 22$~min). A possible infrared counterpart was reported as part of the discovery, though the source is not well resolved and its association with the radio source is unconfirmed. This infrared source has a magnitude of $K_s = 19.73\pm0.28$, consistent with a main sequence star of spectral type between mid-K and mid-M at the estimated distance of $5.7\pm2.9$~kpc. More recently, motivated by additional radio observations showing that the intervals of radio emission are spaced by 8.75~h, \citet{Horvath2025} suggested that the observed 22~min period is the beat period of a binary system, where the compact object has a rotation period of $1265.2197\pm0.0002$~s ($\approx 21$~min), and 8.75~h is the orbital period. The possibility that this could be a binary system with the previously detected infrared source being the companion to a white dwarf, in a similar scenario to the other LPTs confirmed to host white dwarf binaries, motivated this work.

\section{Observations and reduction}

\targ\ was observed with HiPERCAM \citep{hipercam} mounted on the 10.4~m Gran Telescopio Canarias (GTC) during Guaranteed Time Observations (GTO) on the night starting 2024 August 03. Because of its dichroic beamsplitters, HiPERCAM obtains simultaneous data in five filters: $u_s$, $g_s$, $r_s$, $i_s$ and $z_s$, which have similar bandpasses to the traditional $ugriz$ filters of the Sloan Digital Sky Survey, but with improved efficiency. The exposure time was set to 15~sec for the $r_s$, $i_s$ and $z_s$ bands, 30~sec for $g_s$ and 45~sec for $u_s$, with negligible dead-time between exposures due to the frame transfer capabilities of the instrument. The maximum exposure time was defined to enable the detection of periods similar to the spin period of the binary white dwarf pulsar AR~Sco (1.95~min; \citealt{Marsh2016}), taking into account that the origin of the observed 22-min radio periodicity was then unknown and not necessarily associated with the spin of a putative white dwarf. The target was observed for a total of two hours, starting on UTC time 2024-08-03T23:20:25 and finishing on 2024-08-04T01:19:20.69. Conditions were clear with a stable seeing around 0.6--0.7~arcsec.

\begin{figure*}
	\includegraphics[width=\textwidth]{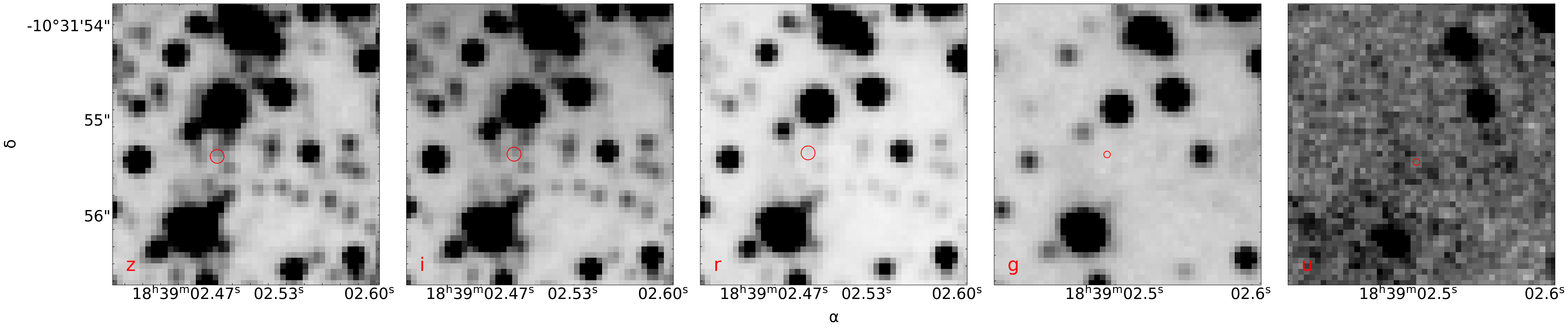}
    \caption{Zoom in on the field around the target for the co-added images for the full run in each passband ($z_s$, $i_z$, $r_s$, $g_s$ and $u_s$  from left to right). The red circle shows the position of the target, with the size of the circle indicating the one-sigma positional uncertainty taking into account both the uncertainty in the radio coordinates and the uncertainty in the astrometric fit of the image (which is not the same for the different filters). The exact coordinates vary slightly ($< 1$'') between filters as the WCS solution is fit independently; $\alpha$ and $\delta$ values shown are for the $z_s$ band.}
    \label{fig:images}
\end{figure*}

The data were reduced using the HiPERCAM pipeline \footnote{\url{https://cygnus.astro.warwick.ac.uk/phsaap/hipercam/docs/html/}}. Frames were bias and flat-field corrected. Fringe correction was applied to the $z_s$ observations. Before performing photometry, it was necessary to determine the expected position of \targ, as its faintness precludes secure visual identification in the optical images. Therefore, we calculated astrometric solutions for the images in each passband using as reference an image created from averaging the first five observed frames, to minimise the impact of jittering. We identified ten relatively bright and isolated stars spread across the CCD and determined their pixel positions in each bandpass through a Gaussian fit. Their right ascension and declination were obtained from the values reported by {\it Gaia} data release 3 (DR3; \citealt{GAIADR3}), taking into account proper motions to obtain positions at the time of observing. An astrometric solution was then calculated using {\tt astropy}'s {\tt wcs} package. Using this World Coordinate System (WCS) solution, we calculated the pixel positions of \targ\ in each bandpass using the radio coordinates reported by \citet{Hurley-Walker2023}. Fig.~\ref{fig:images} shows the obtained location of the target in each bandpass.

The derived pixel locations of \targ\ were used to perform photometry. A bright isolated star was used as reference to calculate centroiding corrections to the target's aperture for each individual frame, to account for imperfect tracking and guiding. To minimise contamination caused by field crowding, we carried out point-spread function (PSF) photometry for all bands except $u_s$, for which a good PSF fit could not be obtained, likely due to the low number of bright stars. When doing PSF fitting, a Moffat profile was employed and its shape was determined for each individual frame by fitting ten relatively isolated stars near the location of \targ. For the $u_s$ frames we carried out aperture photometry with an aperture radius set to be 1.5 times the seeing (determined from a Gaussian fit to the reference stars in each frame). We calibrated the photometry using {\tt cam$\_$cal}\footnote{\url{https://github.com/Alex-J-Brown/cam_cal}}, which calculates a zero-point from observations of a standard star and atmospheric extinction corrections using reference stars in the same field as the target. No standard star was observed on the same night, so we used observations of the standard star G93-48 taken under photometric conditions five nights prior. Results within uncertainties were obtained when a star in the same field was used as standard instead, but no stars in the field have available $u$-band magnitude, therefore we opt for using the calibration based on G93-48.

\section{Results}

\subsection{Light curve and magnitude limits}

\begin{figure*}
    \includegraphics[width=0.75\textwidth]{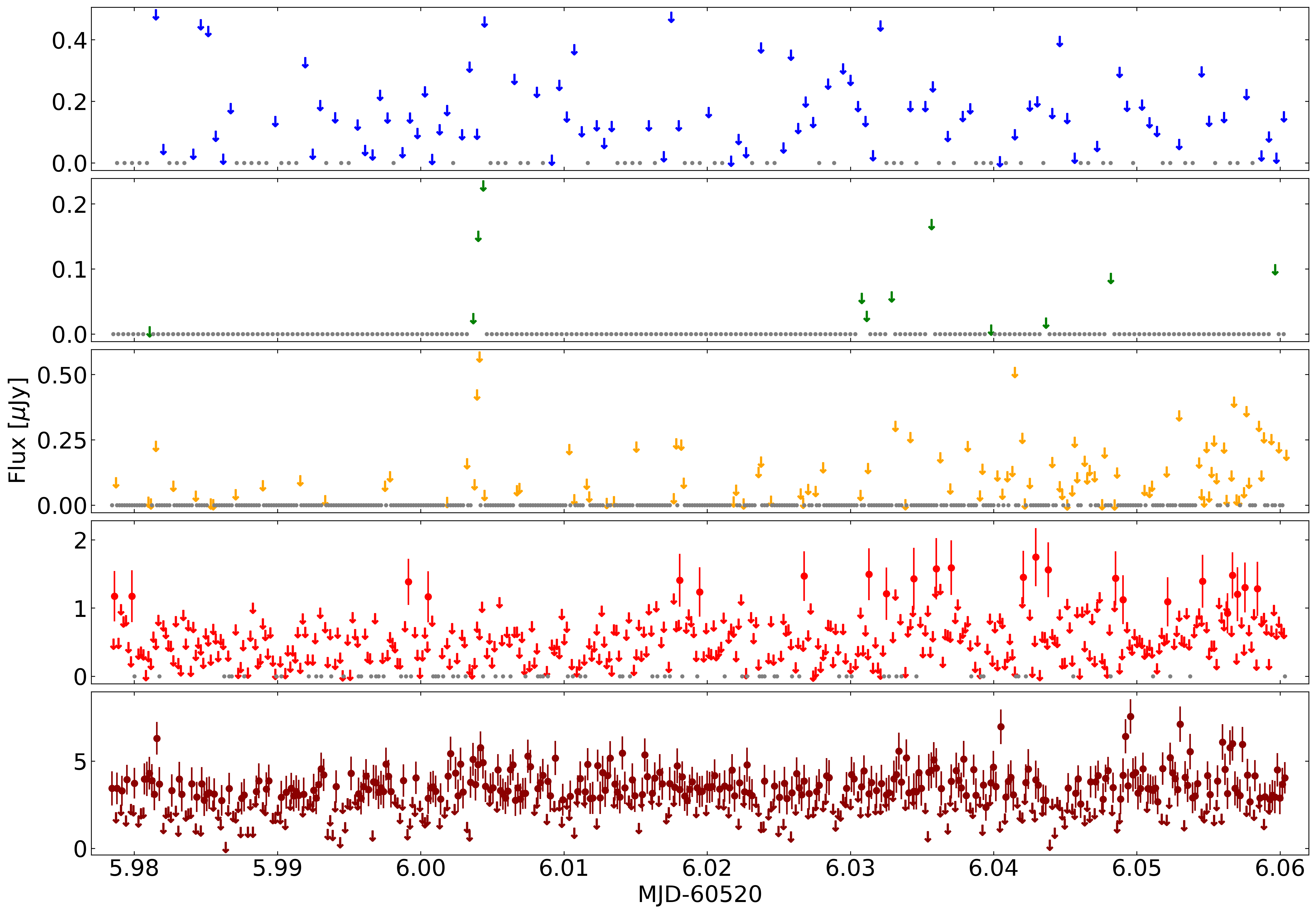}
    \caption{Light curve in the $u_s$, $g_s$, $r_s$, $i_s$ and $z_s$ bands from top to bottom. Symbols with errorbars are measurements were the flux value is significant at a 3-sigma level, downward arrows are values below a 3-sigma significance, and grey dots are used when the target was not detected above the background. \targ\ was only significantly detected in $z_s$.}
    \label{fig:lc}
\end{figure*}

Figure~\ref{fig:lc} shows the calibrated light curves for \targ. There is significant flux in the $z_s$ band, only marginal detection in the $i_s$ band, and no clear detection in the $u_s$, $g_s$ and $r_s$ bands. Based on these light curves, we estimated the fluxes and magnitudes reported in Table~\ref{tab:fluxes}.

\begin{table}
	\centering
	\caption{Derived fluxes and magnitudes of \targ\ in the five HiPERCAM filters. Limits represent the 3-$\sigma$ confidence level.}
	\label{tab:fluxes}
	\begin{tabular}{ccc} 
		\hline
		Band & Flux [$\mu$Jy] & AB magnitude \\
		\hline
		$z_s$ & $3.0\pm1.1$ & $22.70\pm0.53$ \\
		$i_s$ & $<1.6$ & $>23.4$ \\
		$r_s$ & $<0.47$ & $>24.5$ \\
            $g_s$ & $<0.18$ & $>25.5$ \\
            $u_s$ & $<0.48$ & $>24.7$ \\
		\hline
	\end{tabular}
\end{table}

\subsection{Period search}
\label{sec:periods}

\begin{figure}
\centering
    \includegraphics[width=0.45\textwidth]{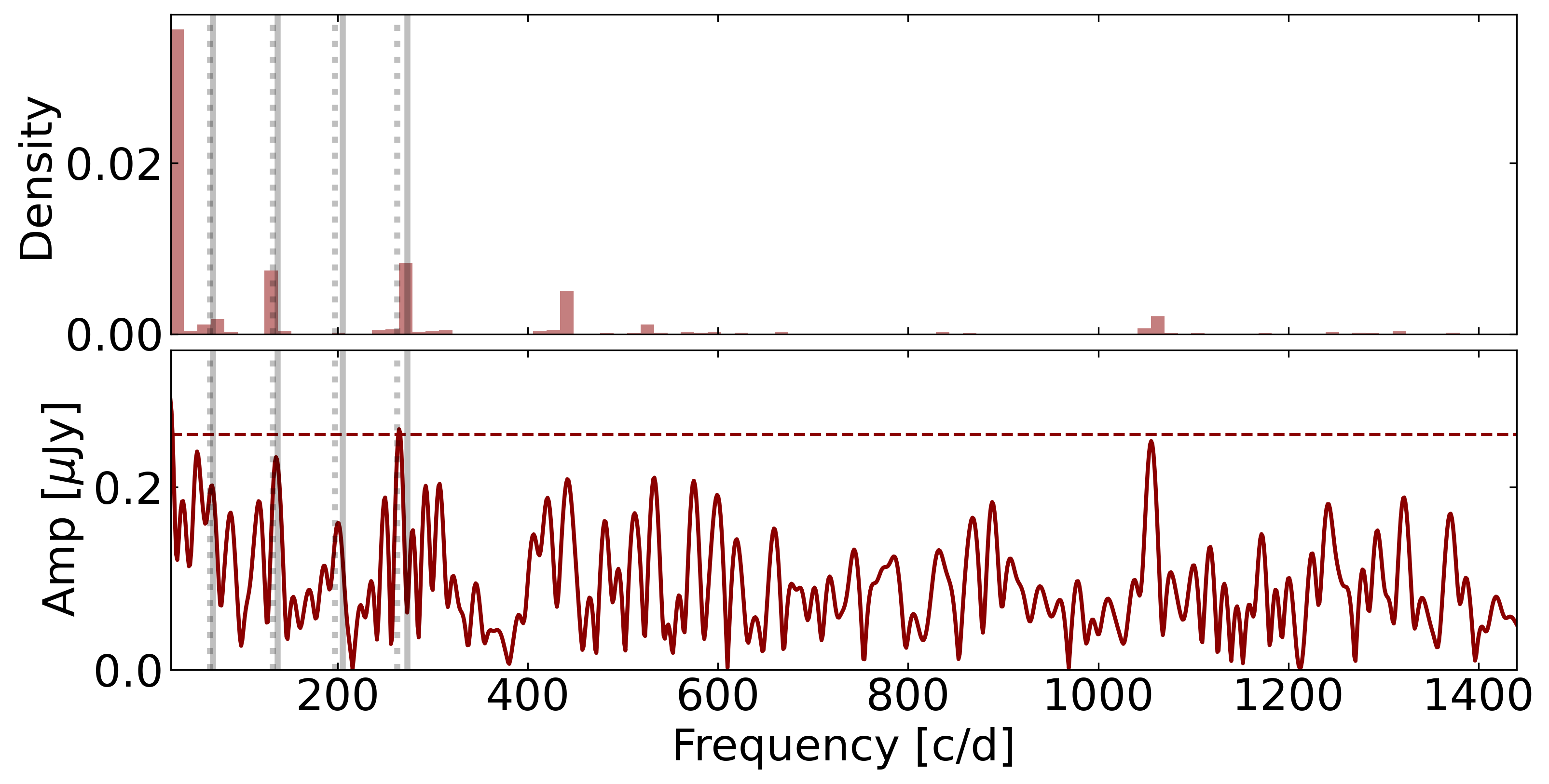}
    \caption{The bottom panel shows the Fourier transform for the $z_s$ light curve, with a 3-$\sigma$ detection threshold indicated by the dashed horizontal line. The top panel shows the distribution of periods obtained from phase-dispersion minimisation. The dashed vertical grey lines in each plot correspond to the period reported by \citet{Hurley-Walker2023}, as well as its first to third harmonics, and solid lines are the same for the spin period proposed by \citet{Horvath2025}.}
    \label{fig:ft}
\end{figure}

To search for periodic behaviour, we calculated the Fourier transform of the $z_s$ light curve, where the target is detected. A 3-$\sigma$ peak-detection threshold was calculated via Monte Carlo, by shuffling the timestamps five thousand times to remove any underlying periodic signal and recalculating the Fourier transform. The maximum amplitude was recorded each time, and our detection threshold was set to the 99.7 per cent quantile of the distribution of maxima. To account for non-sinusoidal signals, we also carried out phase-dispersion minimisation. Flux uncertainties were taken into account with a Monte Carlo approach. We repeated the phase-dispersion minimisation process five thousands times, each time sampling the fluxes assuming a normal distribution with standard deviation given by the flux errors. We recorded the frequency of minimum dispersion for each iteration. Results from both approaches are shown in Fig.~\ref{fig:ft}. There is one peak above the detection threshold in the Fourier transform, and it is consistent with the third harmonic of \targ's reported radio periods. Marginal peaks consistent with other harmonics are also seen, in particular the first harmonic (with a significance of $2.6\sigma$). These same peaks are prominent in the phase-dispersion minimisation, though outranked by low-frequency noise. 

\begin{figure}
\centering
    \includegraphics[width=0.4\textwidth]{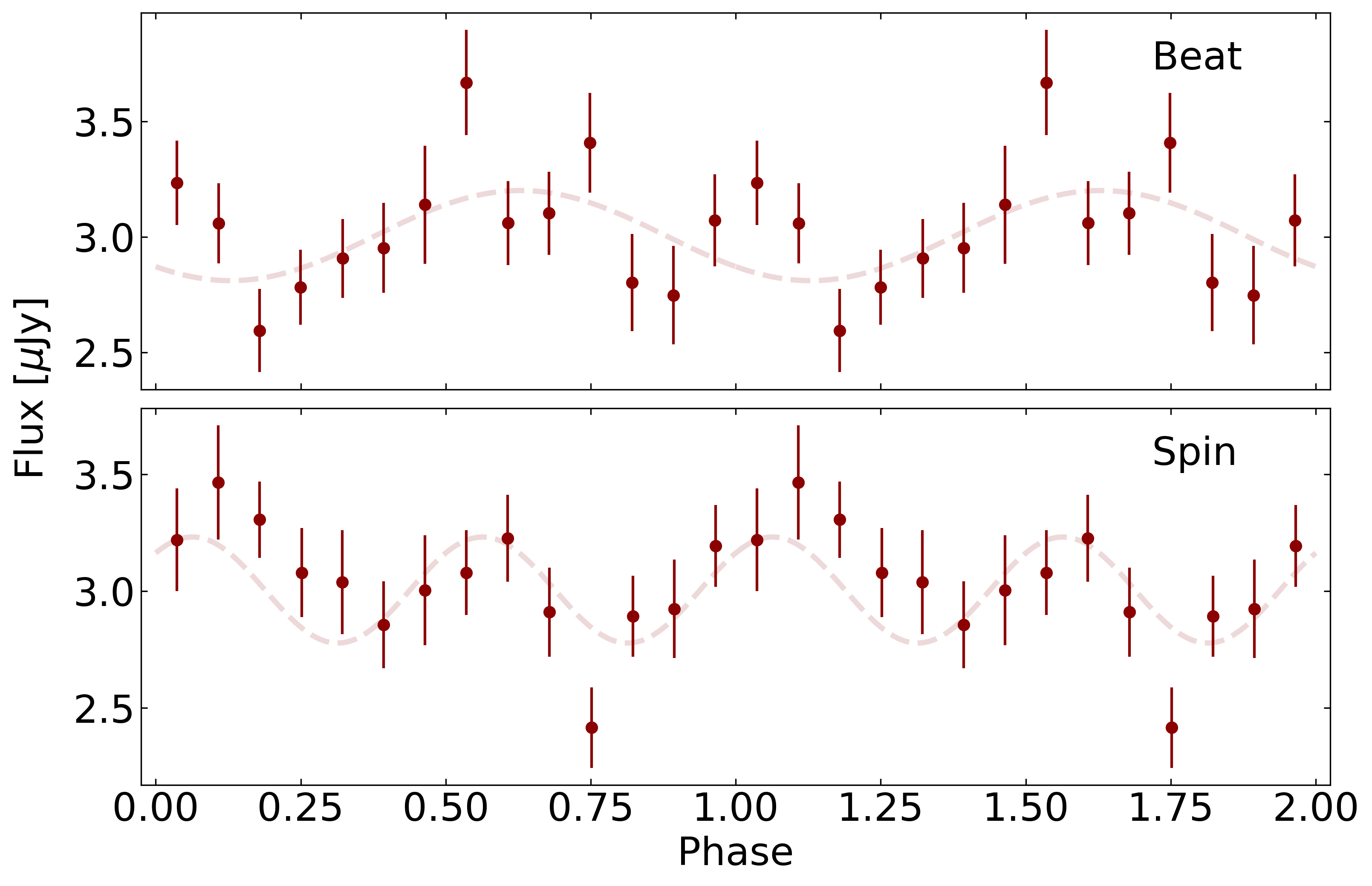}
    \caption{Folded light curves for the $z_s$ observations of \targ, with the data binned to 15 phase bins for clarity. The top panels show the data folded on the beat period, and the bottom panels shows the same for the spin period. The dashed lines show the sinusoidal fits that are favoured over the mean by an F-test at a 95\% confidence level.}
    \label{fig:phase}
\end{figure}

We also folded the light curves to the period of $1318.1957(2)$~sec (presumably the beat), reported by \citet{Hurley-Walker2023}, as well as the reported spin of $1265.2197\pm0.0002$~s from \citet{Horvath2025}. The $T_0$ was set to an arbitrary value of 60520 in all cases, as no ephemeris is published. To assess the significance of any observed behaviour, we fit the phase curves with a sine function and performed an $F$-test with the null hypothesis that the data are well described by simply the mean flux. The null hypothesis is rejected at a chosen 95 per cent confidence level when the data are folded on the beat period ($p$-value = 0.013). For the spin, the same is true if the data are fitted with the first harmonic rather than the fundamental period ($p$-value = 0.048). Phase-folded light curves are shown in Fig.~\ref{fig:phase}. We also performed a different test in a Bayesian framework and compared the Bayesian information criterion (BIC) of the constant and sine models. In both cases, we allowed for a fudge factor in the fit that would inflate the uncertainties, to simulate a situation where uncertainties are injected into the model. The BIC for the constant model is 607, whereas the BIC for the sine model is 615. Therefore, the BIC values are close enough that we cannot prefer a model over the other in this case.

\section{Discussion}

The likely far distance of \targ\ ($5.7\pm2.9$~kpc), combined with its location near the Galactic plane ($b \approx -2^{\circ}$), implies potentially very high reddening of up to $A_V \approx 6$ \citep{Schlafly2011}. Therefore, the lack of detection in the bluer bands is not particularly constraining for a white dwarf scenario, as even fairly hot white dwarfs would lie below our detection limits under these conditions. Taking our derived upper limits and fluxes, as well as the $K_s$ magnitude reported by \citet{Hurley-Walker2023}, we calculated the maximum white dwarf temperature that would remain undetectable as a function of distance and $A_V$, considering a wide range of values given the very uncertain distance of \targ. 

We used models calculated by \citet{Koester2010}, available through the Spanish Virtual Observatory (SVO)\footnote{\url{https://svo2.cab.inta-csic.es/theory/newov2/index.php?models=koester2}}, which span effective temperatures between 6\,000 and 80\,000~K. The $\log~g$ has little impact on broadband fluxes and was fixed at 8.0, near the mean value for white dwarfs \citep[e.g][]{Kepler2021}. The radius was fixed at a representative value of 0.0135~R$_{\odot}$ regardless of temperature (it varies by only $\approx \pm 5$ per cent as a white dwarf cools down). A low-mass white dwarf could have a radius significantly larger ($>20$\%) than our assumed value, but these are found to be remarkably rare in the class of interacting white dwarf plus red dwarf binaries \citep{Zorotovic2011, Pala2020} to which \targ\ is likely related, therefore we consider our assumed radius to be a suitable upper limit to verify if our observations can exclude the presence of a white dwarf in \targ.

As towards the infrared a red dwarf companion could contribute significantly or even dominate the flux, we also do separate calculations taking its contribution into account. For the red dwarf, we used NextGen solar metallicity models \citep{Allard1997}, also available from SVO\footnote{\url{https://svo2.cab.inta-csic.es/theory/newov2/index.php?models=NextGen}}. We assumed $\log~g = 4.5$ and used two different temperatures, 3000~K and 4000~K, to probe different regimes. The radius was fixed at a typical value of $0.2$~R$_{\odot}$. Extinction was applied using the python {\tt extinction} module\footnote{\url{https://extinction.readthedocs.io/en/latest/}} assuming the law of \citet{FM2007} that uses $R_V = 3.1$.

\begin{figure}
    \centering
    \includegraphics[width=0.8\columnwidth]{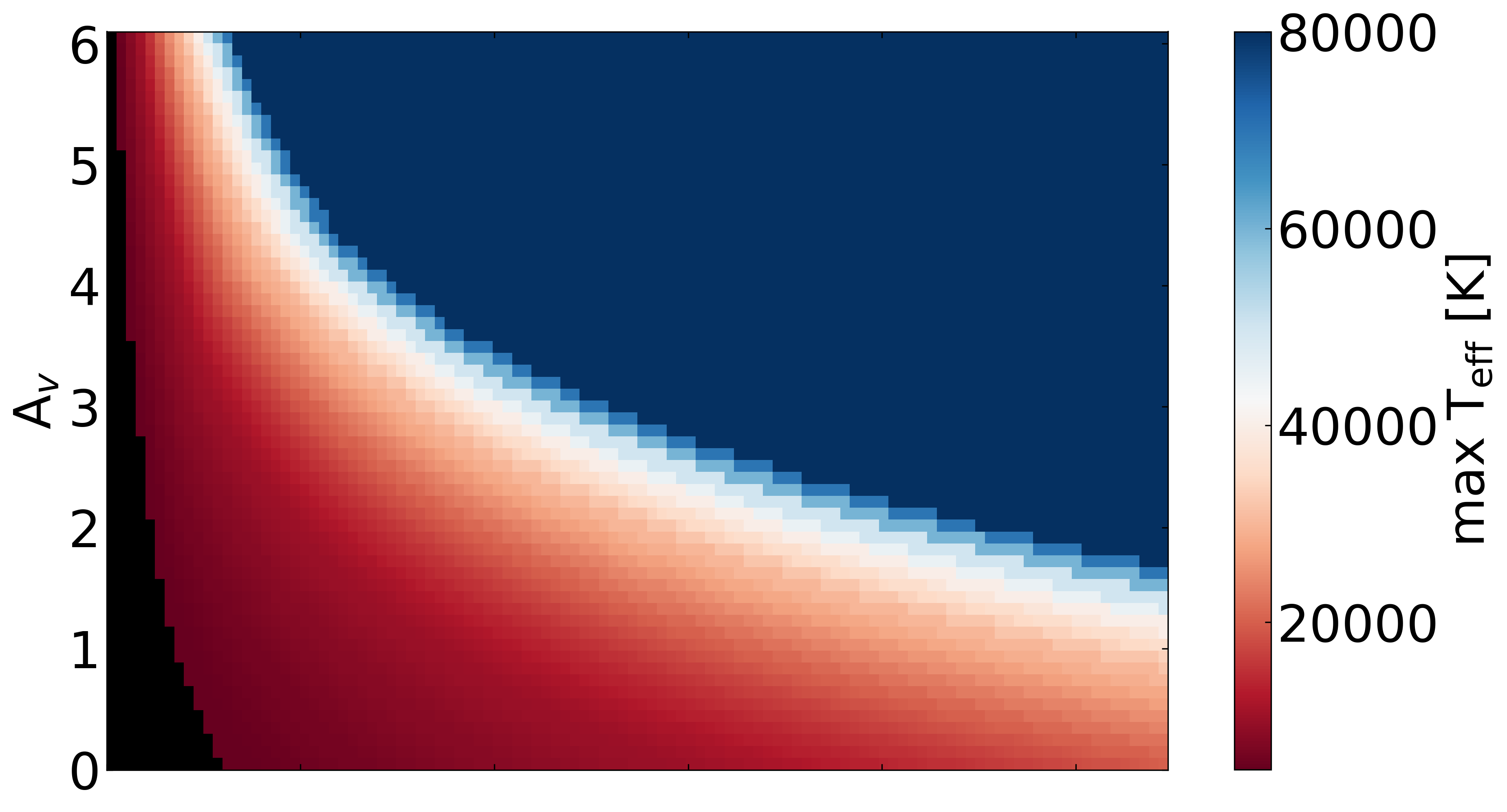}
	\includegraphics[width=0.8\columnwidth]{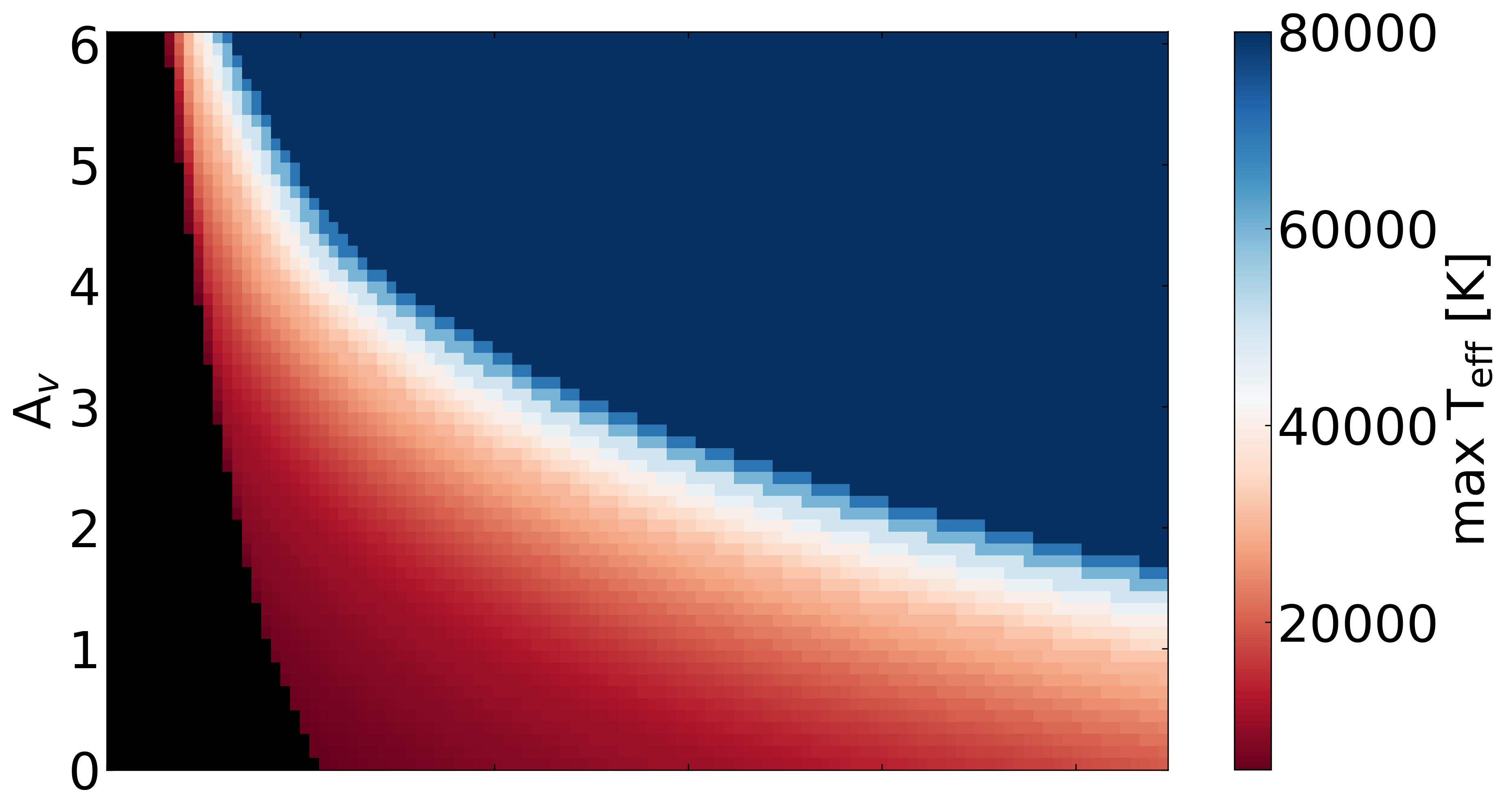}
	\includegraphics[width=0.8\columnwidth]{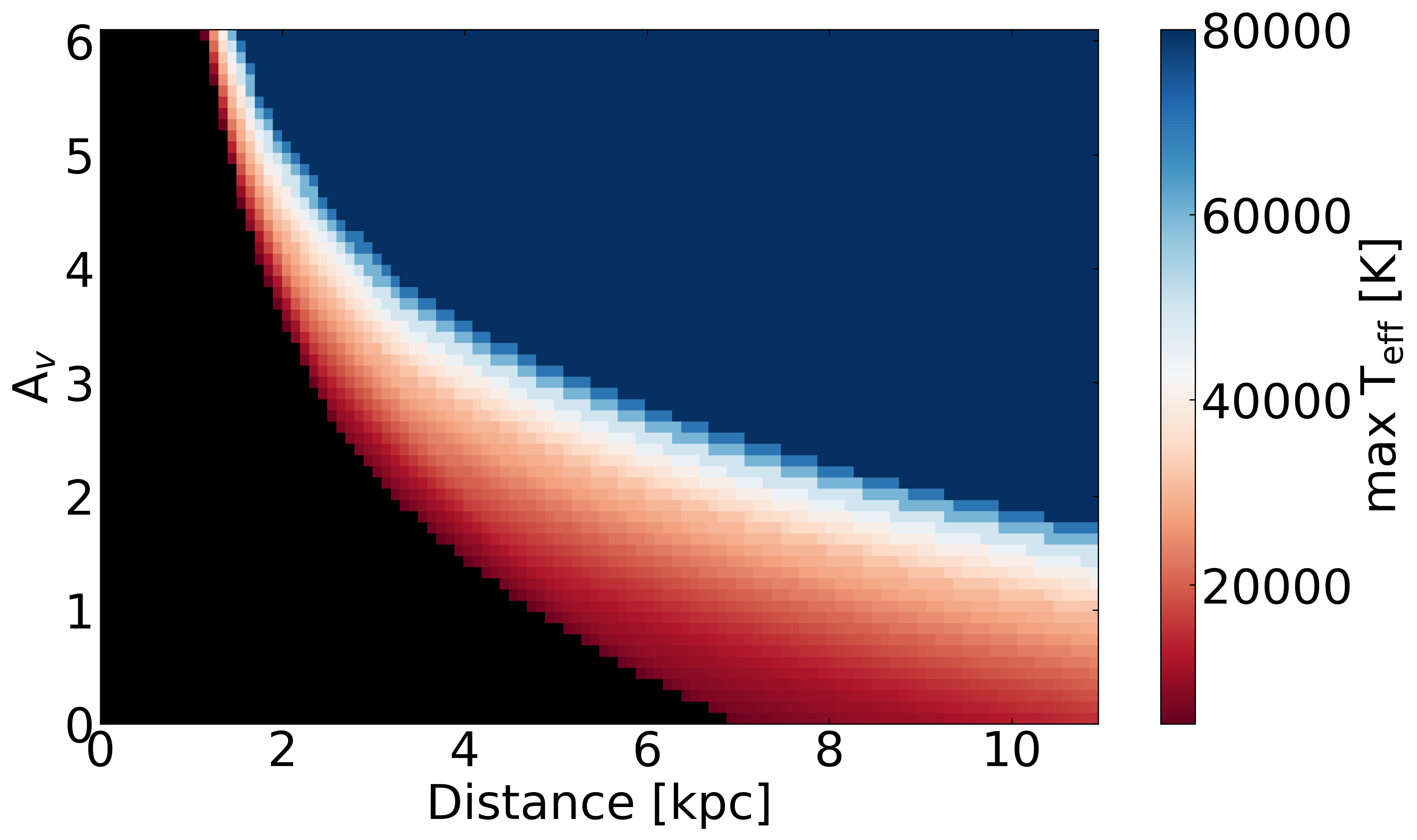}
    \caption{Maximum temperature of a white dwarf that would not be detected as a function of $A_V$ and distance. The top panel shows limits for a white dwarf only, the middle panel assumes a companion with temperature 3000~K, and the bottom panel assumes 4000~K. The maximum allowed white dwarf temperature is indicated by the colourbar; black is used when any system would be detected.}
    \label{fig:maxT}
\end{figure}

Fig.~\ref{fig:maxT} shows the results. A single white dwarf can only be fully excluded for distances closer than 100~pc. When the contribution of a companion is taken into account, we can rule out distances closer than 600~pc, as this would result in a detection considering our magnitude limits. Beyond these distances, limits highly depend on the distance and reddening. For the range 2.8-8.6~kpc \citep[one-sigma interval of the distance reported by][]{Hurley-Walker2023}, even an 80\,000~K white dwarf would remain undetectable in all cases unless the $A_V$ is much smaller than the total line of sight reddening of 6. These limits would become less restrictive if the white dwarf is smaller than the canonical size (i.e. if it is a massive white dwarf). Rerunning this experiment with a 0.005~R$_{\odot}$ white dwarf, we find that a single massive white cannot be excluded, unless both the reddening and distance are much smaller than estimated. For the case of a binary the difference is less striking, and essentially only the maximum temperature of a white dwarf that would not be detected is affected: a massive white dwarf would remain undetected at close distances even for very high temperatures.

\begin{figure}
    \centering
	\includegraphics[width=0.9\columnwidth]{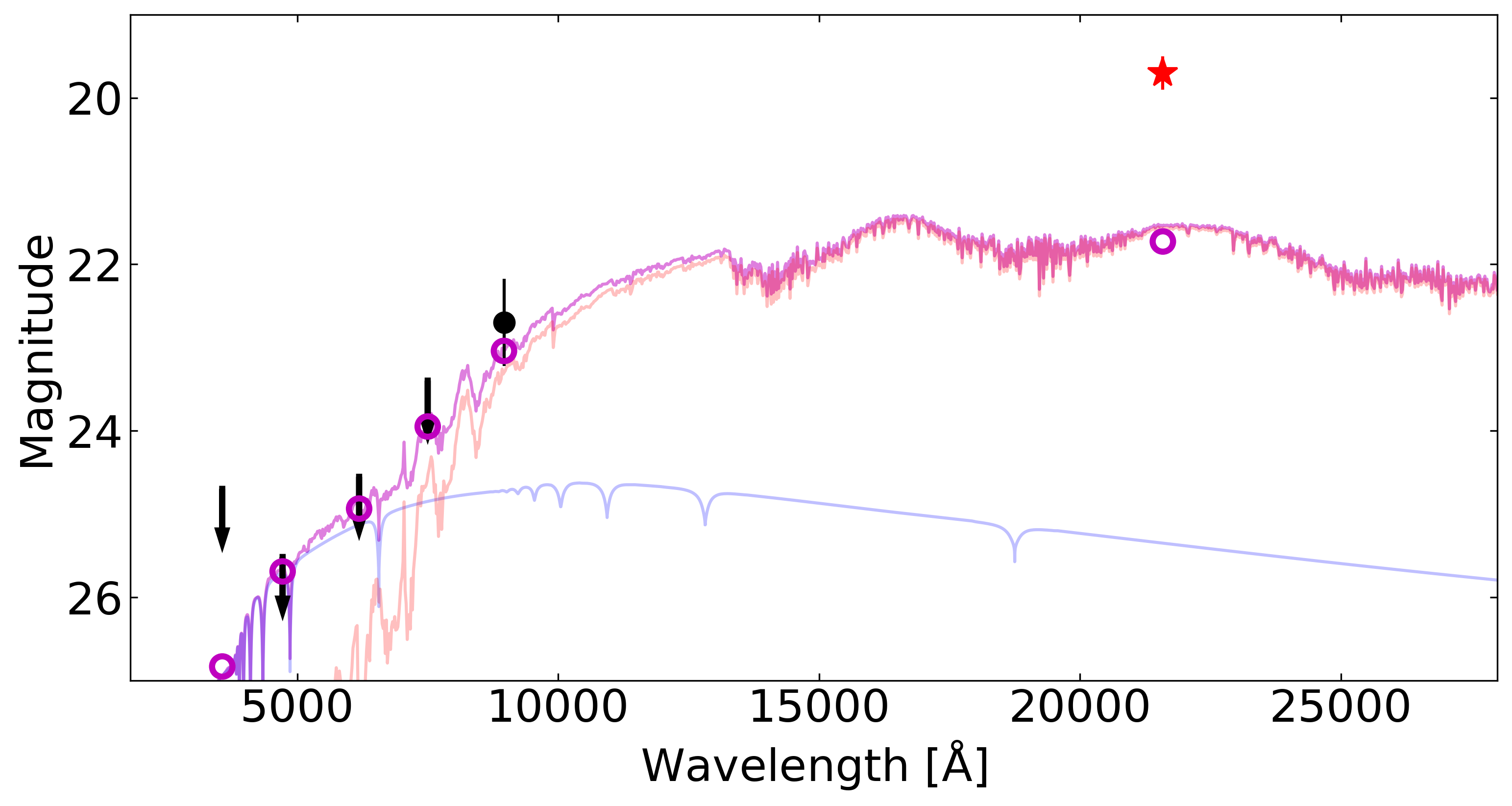}
    \caption{Example SED comparison used to place limits on the maximum effective temperature allowed for a white dwarf in \targ. Magnitudes obtained as part of this work are shown in black, with arrows used for limits, and the red star is the value reported by \citet{Hurley-Walker2023}. The blue line is the white dwarf model and the red line the M-dwarf model, both of which have been reddened and rescaled assuming values as described in the text. The magenta line is the combined flux from both stars. As there is an expected component of non-thermal emission, the observed magnitudes can be brighter than the sum of the stellar components (as is the case for the $K_s$ measurement), but observed values or limits fainter than the model would indicate that a system with a white dwarf of that temperature can be excluded as it would lead to a (stronger) detection. For the case shown in this figure, a white dwarf temperature of 11\,500~K is allowed for a 3000~K companion, 2~kpc distance and $A_V = 2.2$; higher temperatures would cause the expected $g_s$ magnitude to be brighter than our 3-$\sigma$ limit.}
    \label{fig:sed}
\end{figure}

An example spectral energy distribution (SED) is shown in Fig.~\ref{fig:sed}, for a distance of 2~kpc and $A_V = 2.2$, in which case the maximum white dwarf temperature allowed with a 3000~K companion is 11\,500~K; temperatures above this value would result in a significant detection in the $g_s$ band. That is the same temperature as the white dwarfs in the binary white dwarf pulsar systems AR~Sco and J191213.72-441045.1 \citep{Garnavich2021, Pelisoli2024a}, whereas for ILT~J1101+5521 and GLEAM-X~J0704–37 the estimated temperatures are $4500-7500$~K and $7320^{+800}_{-900}$~K, respectively \citep{deRuiter2025, Rodriguez2025}. Temperatures in the range of the value found for these LPTs cannot be excluded in the binary scenario for distances larger than 4.0~kpc or $A_V$ larger than 1.8, which is likely the case for \targ.

As our data do not rule out a white dwarf in \targ, the fact that our highest signal-to-noise ratio light curve (the $z_s$ band) shows evidence for periodic behaviour, as found in Section~\ref{sec:periods}, is encouraging for a white dwarf scenario. As discussed in the introduction, this would likely mean that the system is a binary where particles are fed into the white magnetosphere by a red dwarf companion \citep[which is the proposed infrared counterpart,][]{Hurley-Walker2023}. A period of $\approx 20$~min would be too short to be the orbital period of such a binary, as the red dwarf would be overfilling its Roche lobe by a factor of $\approx 2$. The minimum period for such a binary is around 80~minutes \citep{Knigge2011}. Instead, the observed period would be associated with the rotation period of the white dwarf, as observed for the binary white dwarf pulsars and as proposed by \citet{Horvath2025} for \targ. A 21~min spin period is quite typical for a white dwarf that has accreted material from a companion \citep[e.g.][]{Hellier1996}, as is believed to be the case for binary white dwarf pulsars \citep{Schreiber2021}. AR~Sco, whose current spin period is 1.95~min, is in a stage of fast spin-down \citep[$P/\dot{P} = 5.6 \times 10^6$~yr,][]{Pelisoli2022}, implying that in the future it may show a spin similar to \targ, which could therefore be in a more advanced evolutionary stage than AR Sco. It is worth noting that there is an upper limit to \targ's spin down of $\dot{P} < 3.6 \times 10^{-13}$~s~s$^{-1}$ \citep{Hurley-Walker2023}, which excludes a spin-down matching AR~Sco's \citep[$6.62\pm0.11 \times 10^{-13}$~s~s$^{-1}$,][]{Pelisoli2022}, but only by a small factor that could perhaps be explained by evolution.

One should of course be aware of the possibility of a chance alignment and take into consideration that the detected optical/infrared source might not be associated with the radio source. Using {\tt sep} \citep{sextractor, Barbary2016}, we find that our $2.8\times1.4$~arcmin $z_s$ HiPERCAM field contains around 2000 sources, resulting in a source density of 0.14~arcsec$^{-2}$. The positional uncertainty, taking into account the uncertainties on the radio coordinates and on the astrometric solution of the $z_s$ image, is 0.4~arcsec. Therefore, given the source density, the chance of a source being within one-sigma of the radio coordinates by chance is around 7 per cent. This is non-negligible, though the marginally sinusoidal behaviours shown by the $z_s$ light curve when folded to the radio periods would be hard to explain in this case.

Although our findings are consistent with the white dwarf scenario, as we have no clear direct detection of a white dwarf in the system and there is a non-negligible probability of chance alignment with the source detected in (at least) $z_s$, we cannot fully rule out a neutron star scenario. The main explanation for LPTs when considering a neutron star origin are magnetars. However, no X-ray emission is detected at the position of \targ, and implied limits are orders of magnitudes below those measured for magnetars \citep{Hurley-Walker2023}, making this possibility unlikely. The double neutron star scenario proposed by \citet{Turolla2005}, where the radio is explained by a shock from the interaction between the wind of one pulsar with the magnetosphere of the companion, remains a possibility. The precessing pulsar model of \citet{Zhu2006} is another possibility, although in this case the lack of a detected short spin period for the pulsar, despite there being observations with time resolution of a few tens of $\mu$s, is an argument against it.



\section{Summary \& Conclusions}

We carried out follow-up observations of the LPT \targ\ \citep{Hurley-Walker2023, Horvath2025} using HiPERCAM to probe the existence of a white dwarf source at the position of the radio coordinates. We were motivated by the detection of an infrared counterpart \citep{Hurley-Walker2023} and by the recent work of \citet{Horvath2025}, who have argued that \targ\ can be explained by the same model as the binary white dwarf pulsars \citep{Marsh2016, Pelisoli2023, CastroSegura2025}. We confidently detect a source in the $z_s$ band and derive limits for the other filters (Table~\ref{tab:fluxes}). 

The spin and beat periods reported by \citet{Horvath2025} and \citet{Hurley-Walker2023} are not above a 3$\sigma$ detection threshold in our data, but some of their harmonics show peaks with a significance $>2.5\sigma$ and are also recovered by phase-dispersion minimisation. Additionally, folding our $z_s$ data to the reported periods results in a behaviour that is better explained by a sinusoidal than a constant. All of this points to the existence of periodic signals in our data, as expected if the radio source is a white dwarf system.

Based on the derived magnitudes and limits, we can only fully exclude a white dwarf plus red dwarf binary for distances closer than 600~pc, below the one-sigma interval for \targ\ of $5.7\pm2.9$~kpc. Binary white dwarfs with temperatures similar to what was found for other LPTs cannot be excluded, unless the system is closer than 4.0~kpc or has $A_V < 1.8$, which is unlikely the case for \targ. 

There is a non-negligible chance alignment probability of 7 per cent, therefore we cannot categorically associate the HiPERCAM source with \targ. However, considering that the source seems to show periodic behaviour with the same period as \targ, we consider that there is evidence for the scenario where \targ\ contains a white dwarf with a red dwarf companion, as proposed by \citet{Horvath2025}. In this scenario, the observed radio period of 22~min is the beat period, similar to what is observed for  AR~Sco. Continued monitoring and deeper imaging is required to unambiguously confirm this scenario. Direct detection of the orbital period and/or of the white dwarf would be the confirmation that \targ\ is another LPT containing a white dwarf plus red dwarf system. Additionally, improved distance or coordinate determination would allow us to place stricter constraints using the existing data.

\section*{Acknowledgements}

IP acknowledges support from the Royal Society through a University Research Fellowship (URF\textbackslash R1\textbackslash 231496). This project has received funding from the European Research Council under the European Union’s Horizon 2020 research and innovation programme (Grant agreement numbers 101002408 – MOS100PC). We thank Natasha Hurley-Walker, Nanda Rea, and Csan\'{a}d Horv\'{a}th for useful discussions on the nature and observed periods of \targ.


\section*{Data Availability}

All data analysed in this work can be made available upon reasonable request to the authors.



\bibliographystyle{mnras}
\bibliography{gpmj1839-10} 





\bsp	
\label{lastpage}
\end{document}